\documentclass[aps,pra,a4paper,preprint]{revtex4-1}
\usepackage{amsmath}
\usepackage{amssymb}
\usepackage{bm}
\usepackage{xcolor}
\usepackage{graphics}
\usepackage{graphicx}

\begin{document}
\title{Analytical expressions of non-relativistic static polarizabilities for hydrogen-like ions}
\author {Xuesong Mei $^{1,3}$}
\author {Wanping Zhou $^{2}$}
\author {Haoxue Qiao $^{1}$} \thanks{Email address: qhx@whu.edu.cn}

\affiliation {$^1$School of Physics and Technology, Wuhan University, Wuhan 430072, China}
\affiliation {$^2$Engineering and Technology College, Hubei University of Technology, Wuhan 430000, China}
\affiliation {$^3$State Key Laboratory of Magnetic Resonance and Atomic and Molecular Physics, Wuhan Institute of Physics and Mathematics, Chinese Academy of Sciences, Wuhan 430071, China}

\begin{abstract}
In this work, analytical formulas for the static multipole polarizabilities of hydrogen-like ions are derived by using the
analytical wave functions and the reduced Green function and by applying a numerical fitting procedure.
Our results are then applied to the studies of
blackbody radiation shifts to atomic energy levels at different temperatures.
Our analytical results can be served as a benchmark for other theoretical methods.
\end{abstract}

\maketitle

\section{Introduction}
Hydrogen is one of the most studied atomic systems. Hydrogen is also a prototype for studying alkali atoms and highly charged hydrogen-like ions.
The static polarizabilities of hydrogen-like systems are very important for high-precision spectroscopy experiments. Different computational methods
have been developed for evaluating static polarizabilities of simple atomic systems~\cite{PhysRevA.62.052502, PhysRevA.68.044503, cohen2006numerical, PhysRevA.86.012505}. The analytical expressions for both the Schr{\"o}dinger- and Dirac-Green functions of hydrogen have been derived by
Swainson and Drake~\cite{Swainson_1991a,Swainson_1991b,Swainson_1991c}. In their approach, the wave functions and the Green functions are expressed as infinite series involving Laguerre functions, which is a Sturmian form in essence. A formal scheme for deriving the dipole polarizabilities of hydrogen has been given by Krylovetsky \textit{et al.}~\cite{krylovetsky2001generalized}, where the Sturmian polynomials are also used in their work to describe the wave functions and the Green function. Analytical expressions for the dipole polarizabilities of hydrogen in an arbitrary atomic state have been obtained by Baye~\cite{PhysRevA.86.062514}. However, analytical results for multipole polarizabilities of hydrogen are still less investigated, which will be the main focus of the present work.
This paper is organized as follows. In Sec.~\ref{Sec.Method}, we introduce the theoretical formalism. In Sec.~\ref{sect_scalar_pol}, we demonstrate our fitting procedure for obtaining analytical expressions of static multipole polarizabilities. In Sec.~\ref{BBR-shift}, we apply our results to the calculations of the blackbody radiation shifts of the hydrogen atom.

\section{Formalism}\label{Sec.Method}
In spherical coordinates, the wave function for a spherically symmetric potential $V(r)$ can be written in the form
\begin{equation}
\begin{aligned}
\psi_{nlm}(r,\theta,\phi)=R_{nl}(r)Y_{lm}(\theta,\phi)\,,
\end{aligned}
\end{equation}
where $R_{nl}(r)$ is the radial part and $Y_{lm}(\theta,\phi)$ is the spherical harmonics. In atomic units, the radial equation for $R_{nl}(r)$ is
\begin{equation}
\begin{aligned}
\left\{
\frac{1}{r^2} \frac{\mathrm{d}}{\mathrm{d} r}
\left(r^2 \frac{\mathrm{d}}{\mathrm{d} r}\right)
-
\frac{l(l+1)}{r^2}
+2[E+V(r)]
\right\} R_{nl}(r)=0\,.
\end{aligned}
\end{equation}
For hydrogen, $V(r)=-1/r$ and the energy eigenvalue is $-1/(2n^2)$.

When an external electric field exists, atomic energy levels will be shifted due to the Stark effect. We can treat the electric field as a perturbation if it is weak. The energy shift $\Delta E$ can then be expanded in terms of the external field $\mathcal{E}$ according to Ref.~\cite{buckingham1967permanent}
\begin{equation}\label{energy_shift_general}
\begin{aligned}
\Delta E =
\bigg[-\frac{1}{2}\alpha_d \mathcal{E}^2
-\frac{1}{6} \alpha_q (\partial\mathcal{E})^2
+\cdots\bigg]
-\frac{1}{3!} \beta \mathcal{E}^3
-\frac{1}{4!} \gamma \mathcal{E}^4-\cdots,
\end{aligned}
\end{equation}
where $\alpha_d$ and $\alpha_q$ are, respectively, the electric dipole and quadrupole polarizabilities, $\partial \mathcal{E}$ is the field gradient at the origin, and $\beta$ and $\gamma$ are the hyperpolarizabilities. In Eq.~\eqref{energy_shift_general}, the terms in square bracket attribute to two-photon process, which means that the electron absorbs one photon and then emits one photon or vice versa.
Considering a multipole electric polarizability, a general expression for a two-photon process can be written as
\begin{equation}\label{def_polarizability}
\begin{aligned}
\alpha^{(\lambda)}_\mu=
\left< \psi \left|
Q^{(\lambda)}_{\mu}
\left(\frac{1}{E_0-H_0}\right)'
Q^{*(\lambda)}_\mu
+
Q^{*(\lambda)}_\mu
\left(
\frac{1}{E_0-H_0}
\right)'
Q^{(\lambda)}_\mu
\right| \psi \right>\,,
\end{aligned}
\end{equation}
where 
$Q^{(\lambda)}_\mu$ is the multipole operator defined by
$${Q}^{(\lambda)}_\mu=\sqrt{4\pi/(2\lambda+1)}r^\lambda Y_{\lambda\mu} (\theta, \phi),
$$
and $\lambda$ is the multiplicity, such as $\lambda=1$ for dipole, $\lambda=2$ for quadrupole and so on. The validity of multipole
expansion of an external field is when the external
field wavelength is much longer than the characteristic length of the atom.
The prime notation on $1/(E_0-H_0)$ indicates that it is the reduced Green function. Rewriting Eq.~\eqref{def_polarizability} in the spatial representation we have
\begin{equation}\label{spatial_representation}
\begin{aligned}
\alpha _{nl}^{(\lambda)\mu}=\int{}\psi _{nlm}^{\dagger}\left( \boldsymbol{r}_1 \right) \left[ \hat{Q}_{\mu}^{(\lambda)}G'\left(\bm{r}_1, \bm{r}_2; E_n \right) \hat{Q}_{\mu}^{*(\lambda)}+\hat{Q}_{\mu}^{*(\lambda)}G'\left( \bm{r}_1, \bm{r}_2; E_n \right) \hat{Q}_{\mu}^{(\lambda)} \right] \psi _{nlm}\left( \boldsymbol{r}_2 \right) \text{d}\boldsymbol{r}_1
\text{d}\bm{r}_2.
\end{aligned}
\end{equation}

Following Refs.~\cite{Swainson_1991a,Swainson_1991c}, the radial part of the non-relativistic wave function can be written as
\begin{equation}\label{Radial_WF}
\begin{aligned}
R_{nl}\left( r \right) =N_{nl}\left( 2r/n \right) ^le^{-r/n} L_{n-l-1}^{2l+1}\left( 2r/n \right)\,,
\end{aligned}
\end{equation}
where $L_{n}^{l}(z)$ is the generalized Laguerre function and $N_{nl}$ is the normalization factor
$$
N_{nl}=\left({2}/{n^2} \right) \sqrt{\left( n-l-1 \right) !/\left( n+l \right) !}\,.
$$
The reduced Green function in spherical coordinates can be expressed as
\begin{equation}\label{GF_general}
\begin{aligned}
G' \left( \boldsymbol{r}_1,\boldsymbol{r}_2;E_n \right) =\sum_{l'm'}{}g_{l'}\left( r_1,r_2;E_n \right) Y_{l'm'}\left( \theta _1,\phi _1 \right) Y_{l'm'}^{*}\left( \theta _2,\phi _2 \right),
\end{aligned}
\end{equation}
where the radial function $g_{l'}(r_1,r_2;E_n)$ can be expressed as a Sturmian polynomial
\begin{equation}\label{Radial_GF}
\begin{aligned}
&
g_{l'}\left( r_1,r_2;E_n \right) =2\left( \frac{2}{n} \right) ^{2l'+1}\left( r_1r_2 \right) ^{l'}e^{-\left( r_1+r_2 \right) /n}
\\&
\times \left\{ \sum_{k=0}^{\infty}{}\frac{k!}{\left( 2l'+1+k \right) !\left( l'+1+k-n' \right)}L_{k}^{2l'+1}\left( \frac{2r_1}{n} \right) L_{k}^{2l'+1}\left( \frac{2r_2}{n} \right) \right.
\\&
+\frac{\left( n-l'-1 \right) !}{2n' \left( n+l' \right) !}\left\{ L_{n-l'-1}^{2l'+1}\left( \frac{2r_1}{n} \right) \left[ \left( n-l' \right) L_{n-l'}^{2l'+1}\left( \frac{2r_2}{n} \right) \right. \right.
\\&
\left. -\left( n+l' \right) L_{n-l'-2}^{2l'+1}\left( \frac{2r_2}{n} \right) \right] +L_{n-l'-1}^{2l'+1}\left( \frac{2r_1}{n} \right) L_{n-l'-1}^{2l'+1}\left( \frac{2r_2}{n} \right)
\\&
\left. \left. +\left[ \left( n-l' \right) L_{n-l'}^{2l'\,\,+1}\left( \frac{2r_1}{n} \right) -\left( n+l' \right) L_{n-l'-2}^{2l'+1}\left( \frac{2r_1}{n} \right) \right] L_{n-l'-1}^{2l'+1}\left( \frac{2r_2}{n} \right) \right\} \right\}\,.
\end{aligned}
\end{equation}
With these expressions, we can rewrite Eq.~\eqref{spatial_representation} in the form of separated radial and angular parts
\begin{equation}\label{radial_part}
\begin{aligned}
\alpha _{nl}^{(\lambda)\mu}&=\frac{8\pi}{2\lambda +1}
\sum_{l'}{}
\left[
\int{}\text{d}r_1\text{d}r_2r_{1}^{2+\lambda}r_{2}^{2+\lambda}R_{nl}\left( r_1 \right) g_{l'}\left( r_1,r_2;E_n \right) R_{nl}\left( r_2 \right)
\right.
\\&
\left.
\times \sum_{m'}{}\int{\text{d}\Omega_1 {\text{d}\Omega _2}}
Y_{\lambda \mu}^{*}\left( \Omega_1 \right)
Y_{lm}^{*}\left( \Omega_1 \right)
Y_{l'm'}\left( \Omega_1 \right)
Y_{\lambda \mu}\left( \Omega_2 \right)
Y_{l'm'}^{*}\left( \Omega_2 \right)
Y_{lm}\left( \Omega_2 \right)
\right]\,.
\end{aligned}
\end{equation}
The treatment of angular part is quite straightforward by using 
\begin{equation}
\begin{aligned}
\int{\text{d}\Omega}
Y_{\lambda \mu} (\Omega)
Y_{lm} (\Omega)
Y_{l'm'} (\Omega)
=\sqrt{\frac{\left( \lambda,l,l' \right)}{4\pi}}\left( \begin{matrix}
	\lambda&		l&		l'\\
	0&		0&		0\\
\end{matrix} \right)
\left( \begin{matrix}
\lambda	& l&		l'\\
\mu &	m&		m'\\
\end{matrix} \right)\,,
\end{aligned}
\end{equation}
where $(\lambda,l,l')\equiv (2\lambda+1)(2l+1)(2l'+1)$. The angular part thus becomes
\begin{equation}
\begin{aligned}
\left( -1 \right) ^{\mu}\frac{\left( \lambda ,l,l' \right)}{4\pi}\left( \begin{matrix}
\lambda&	l&		l'\\
	0&		0&		0\\
\end{matrix} \right) ^2\left( \begin{matrix}
\lambda&	l&		l'\\
\mu&	m&		m'&		\\
\end{matrix} \right) ^2 .
\end{aligned}
\end{equation}
In the case of scalar polarizability, we should sum over magnetic quantum numbers $m$ and $m'$ and average over $m$. Using the relation
$$
\sum_{m_1,m_2}{}\left( \begin{matrix}
	l_1&		l_2&		l_3\\
	m_1&		m_2&		m_3\\
\end{matrix} \right) \left( \begin{matrix}
	l_1&		l_2&		l'_3\\
	m_1&		m_2&		m'_3\\
\end{matrix} \right) =\frac{1}{2 l_3+1}\delta \left( l_3,l'_3 \right) \delta \left( m_3,m'_3 \right)\,,
$$
we can obtain the final expression of the angular part
\begin{equation}\label{angular_part_scalar}
\begin{aligned}
\mathcal{A}^{(\lambda) \mu}_{ll'}=\left( -1 \right) ^{\mu}\frac{\left( 2l'+1 \right)}{4\pi}\left( \begin{matrix}
\lambda&	l&		l'   \\
	0&		0&		0\\
\end{matrix} \right) ^2.
\end{aligned}
\end{equation}
Let us now consider the radial part in Eq.~(\ref{radial_part}). Introducing the following integral
\begin{equation}\label{M_integral}
\begin{aligned}
\mathcal{M}^{(\lambda)}_{nll'}=
\frac{8\pi}{2\lambda +1}|N_{nl}|^2
\int_0^\infty {r_{1}^{2\lambda+1}}R_{nl}\left( r_1 \right) g_{l'}\left( r_1,r_2;E_0 \right) r_{2}^{2\lambda+1}R_{nl}\left( r_2 \right) \text{d}r_1\text{d}r_2\,,
\end{aligned}
\end{equation}
the scalar polarizability can be written in the form
\begin{equation}\label{polarizabilties_splitted}
\begin{aligned}
\alpha^{(\lambda)\mu}_{nl}
=
\sum_{l'}
\mathcal{M}^{(\lambda)}_{nll'}
\mathcal{A}^{(\lambda)\mu}_{ll'}.
\end{aligned}
\end{equation}
It is noted that the range of $l'$ in the above is from $|l-\lambda|$ to $l+\lambda$ with step 2.
Recalling the Sturmian form of the wave function and the reduced Green function Eqs.~\eqref{Radial_WF} and \eqref{Radial_GF}, the radial integral $\mathcal{M}^{(\lambda)}_{nll'}$ can be reduced as a series of the basic integral~\cite{Introduction.to.Special.Functions}
\begin{equation}\label{radial_general_integral}
\begin{aligned}
\int_0^{\infty}{}x^{\rho}e^{-x}L_{\nu}^{\beta}\left( x \right) L_{\nu '}^{\beta '}\left( x \right) dx=\left( -1 \right) ^{\nu +\nu '}\Gamma \left( \rho +1 \right) \sum_k
\binom{\rho -\beta}{\nu -k}
\binom{\rho -\beta'}{\nu'-k}
\binom{\rho +k}{k}\,,
\end{aligned}
\end{equation}
where $\binom{\rho}{k}=\frac{\rho !}{( \rho -k )!k!}$.
Thus $\mathcal{M}^{(\lambda)}_{nll'}$ should be a polynomial of the principal quantum number $n$ and angular quantum number $l$. Therefore, we can assume that the polarizability can be written in the following general form
\begin{equation}\label{alpha_general}
\begin{aligned}
\alpha^{(\lambda)}_s = \sum_{i} C^{(\lambda)}_{i}(l) n^{x_i}\,,
\\
C^{(\lambda)}_{i}(l)=\sum_j D^{(\lambda)}_{ij} l^{y_{ij}}\,,
\end{aligned}
\end{equation}
where $C^{(\lambda)}_i(l)$ and $D^{(\lambda)}_{ij}$ are the expansion coefficients that need to be determined.

\section{Scalar polarizabilities} \label{sect_scalar_pol}
\subsection{Dipole polarizability} \label{dipole_polarizability}
We first discuss the dipole polarizability where $\lambda=1$ and the corresponding dipole operator is $Q^{(1)}_\mu=\sqrt{4\pi/3}~r Y_{1\mu}(\theta,\phi)$. According to Eq.~\eqref{polarizabilties_splitted}, the dipole polarizability is
\begin{equation}
\begin{aligned}
\alpha^{(1)\mu}_{nl}
=
\sum_{l'}
\mathcal{M}^{(1)}_{nll'}
\mathcal{A}^{(1)\mu}_{ll'}\,.
\end{aligned}
\end{equation}
For each $l'$ we fit our expression to an analytical formula and then sum over
$l'$ to obtain a final expression.

We first consider the radial integral
\begin{equation}\label{dipole_radial_integral}
\begin{aligned}
\mathcal{M}^{(1)}_{nll'}=
\frac{8\pi}{3}
|N_{nl}|^2
\int_0^\infty
{r_{1}^{3}}
R_{nl}\left( r_1 \right)
g_{l'}\left( r_1,r_2;E_0 \right)
r_{2}^{3}
R_{nl}\left( r_2 \right) \text{d}r_1\text{d}r_2\,.
\end{aligned}
\end{equation}
According to Eqs.~\eqref{Radial_WF}, \eqref{Radial_GF}, and \eqref{radial_general_integral}, $\mathcal{M}^{(1)}_{nll'}$ in the above equation can be easily evaluated numerically or analytically. It should be noted that, since the binomial coefficient $\binom{m}{n}$ requires both $m$ and $n$ to be non-negative integrals, the range of $k$ in
Eq.~\eqref{Radial_GF} is actually finite. We choose the upper limit of $k$ to be 100, which is sufficient for this work. The radial integral
$\mathcal{M}^{(1)}_{nll'}$ can be written in a similar form as
Eq.~\eqref{alpha_general} but with different coefficients
\begin{equation}\label{radial_general}
\begin{aligned}
\mathcal{M}^{(1)}_{nll'} = \sum_{i} c^{(1)}_{i}(l) n^{x_i},~
c^{(1)}_{i}(l)=\sum_j d^{(1)}_{ij} l^{y_{ij}}.
\end{aligned}
\end{equation}
We set $l$ from $1$ to $20$, and for each of them, we take 20 allowed principal quantum numbers $n$. Since in the dipole case, $l'$ is $|l-1|$ and $l+1$, we have two cases of radial integral $\mathcal{M}^{(1)}_{nl,|l-1|}$ and $\mathcal{M}^{(1)}_{nl,l+1}$. Thus these radial integrals can form a matrix of $20\times 20$ with respect to $n$ and $l$, with their matrix elements denoted by $\mathcal{M}_{nl}^{(1)}(n,l)$. Recalling the simultaneous equations Eq.~\eqref{radial_general}, our task is to determine the coefficients $c_i^{(1)}$ and $d_{ij}^{(1)}$ by a fitting procedure. In order to do this, we must first make some appropriate assumptions. We assume that the power index of $n$ has a pattern of $2\lambda+2i$, $(i=1,2,\cdots,20)$. In the dipole case, it is $4, 6, 8, \cdots$. Then we have an initial form for this radial integral as $c_1^{(1)} n^4 + c_2^{(1)} n^6 + c_3^{(1)} n^8 +\cdots$. The maximum value of $x_i$ is still unknown at this stage and we temporarily set it to be 20. We first fit $c_i^{(1)}$ to the exact values by solving the following equations
\begin{equation}
\begin{aligned}
\mathcal{M}_{nl}^{(1)}(n_j,l_k) = \sum_{i=1}^{20} c_i^{(1)} n_j^{x_i}, \ \ \ \ j, k=1, 2, \cdots,20,
\end{aligned}
\end{equation}
where $\mathcal M_{nl}^{(1)}(n_j, l_k)$ are the matrix elements evaluated numerically according to Eq.~\eqref{dipole_radial_integral}. 
For given $l_k$, we select 20 values of $n_j$ and solve the resulting equations simultaneously. We show the results of $c_i^{(1)}$ in Table~\ref{fitting_table} of Appendix~\ref{Ftable} to demonstrate the fitting process. We can see that only the coefficients $c_1^{(1)}$ and $c_2^{(1)}$ are non-zero in the dipole case. The fitted values of $c_i^{(1)}$ can be written in two $20\times20$ matrices
$$
c^{(1)} \left( l'=l+1 \right) =\left( \begin{matrix}
	\frac{3}{2}&		\frac{21}{4}&		0&		0&		\cdots&		0&		0\\
	\frac{17}{8}&		\frac{127}{8}&		0&		0&		\cdots&		0&		0\\
	\frac{11}{4}&		\frac{139}{4}&		0&		0&		\cdots&		0&		0\\
	\vdots&		\vdots&		\vdots&		\vdots&		\vdots&		\vdots&		\vdots\\
	\frac{97}{8}&		\frac{22703}{8}&		0&		0&		\cdots&		0&		0\\
	\frac{51}{4}&		\frac{13179}{4}&		0&		0&		\cdots&		0&		0\\
	\frac{107}{8}&		\frac{30385}{8}&		0&		0&		\cdots&		0&		0\\
\end{matrix} \right),
\\ \ \
c^{(1)}(l'=|l-1|)=\left( \begin{matrix}
	\frac{1}{4}&		\frac{-1}{4}&		0&		0&		\cdots&		0&		0\\
	-\frac{3}{8}&		\frac{3}{8}&		0&		0&		\cdots&		0&		0\\
	-1&		\frac{1}{4}&		0&		0&		\cdots&		0&		0\\
	\vdots&		\vdots&		\vdots&		\vdots&		\vdots&		\vdots&		\vdots\\
	-\frac{83}{8}&		-\frac{12973}{8}&		0&		0&		\cdots&		0&		0\\
	-11&		-\frac{7759}{4}&		0&		0&		\cdots&		0&		0\\
	-\frac{93}{8}&		-\frac{18375}{8}&		0&		0&		\cdots&		0&		0\\
\end{matrix} \right)\,,
$$
where the row and column index is, respectively, $l$ and $n$. 
After determining $c^{(1)}_i$, we further solve the following equations to obtain $d^{(1)}_{ij}$
\begin{equation}
\begin{aligned}
c^{(1)}(l_i, x_k)=\sum_{j=1}^{20} d^{(1)}_{j} l_i^{y_j}, \ \ \ (i,k=1,2,\cdots 20)\,,
\end{aligned}
\end{equation}
where $c^{(1)}(l_i,x_k)$ are matrix elements. At this stage, we assume that $y_j$ are non-negative integers and $l$ is from 1 to 20, the same as the $c_i^{(1)}$ fitting.
After finishing the above fitting process, we substitute $d_{ij}^{(1)}$ into Eq.~\eqref{radial_general} and obtain the following radial analytical expressions 
\begin{equation}\label{ratial_dipole}
\begin{aligned}
\mathcal{M}^{(1)}_{nl,|l-1|}&=
\frac{8\pi}{3}
\left[
\left(-\frac{3 l^3}{8}+\frac{15 l^2}{8}-\frac{19 l}{8}+\frac{5}{8}\right) n^4+\left(\frac{7}{8}-\frac{5 l}{8}\right) n^6 \right]
,
\\
\mathcal{M}^{(1)}_{nl,l+1}
&=
\frac{8\pi}{3}\left[
\left( \frac{3l^3}{8}+3l^2+\frac{29l}{4}+\frac{21}{4} \right) n^4+\left( \frac{5l}{8}+\frac{3}{2} \right) n^6 \right].
\end{aligned}
\end{equation}
By including the following angular coefficients defined in Eq.~(\ref{angular_part_scalar}) with $\mu=0$
\begin{equation}
\begin{aligned}
\mathcal{A}^{(1)0}_{l,|l-1|}
=
\frac{1}{4\pi}
\frac{l}{2 l+1},~
\mathcal{A}^{(1)0}_{l,l+1}
=
\frac{1}{4\pi}
\frac{l+1}{2 l+1},
\end{aligned}
\end{equation}
we finally obtain the analytical expression for
the scalar dipole polarizability denoted by $\alpha^{(1)}_s$
\begin{equation}\label{eq_alpha1}
\begin{aligned}
\alpha^{(1)}_s
&=
\mathcal{M}^{(1)}_{nl,|l-1|} \mathcal{A}^{(1)0}_{l,|l-1|}
+
\mathcal{M}^{(1)}_{nl,l+1} \mathcal{A}^{(1)0}_{l,l+1}
\\&
=\frac{7}{4} \left(l^2+l+2\right) n^4+ n^6\,,
\end{aligned}
\end{equation}
which agrees with Ref.~\cite{PhysRevA.86.062514}.

\subsection{Multipole polarizabilities} \label{multipole polarizabilies}
The expressions for higher-order polarizabilities can be obtained in a similar way. In this work, we mainly deal with the quadrupole ($\lambda=2$) and octupole ($\lambda=3$) polarizabilities. Again, we focus on scalar polarizabilities.
For the quadrupole case, ${Q}^{(2)}_\mu=\sqrt{4\pi/5}~r^2 Y_{2\mu} (\theta, \phi)$ and the orbital quantum number $l'$ of the intermediate states takes 3 different values: $|l-2|,~l,~l+2$. The radial integral is now
\begin{equation}\label{quadrupole_radial_integral}
\begin{aligned}
\mathcal{M}^{(2)}_{nll'}=
\frac{8\pi}{5}
|N_{nl}|^2
\int_0^\infty
{r_{1}^{5}}
R_{nl}\left( r_1 \right)
g_{l'}\left( r_1,r_2;E_0 \right)
r_{2}^{5}
R_{nl}\left( r_2 \right) \text{d}r_1\text{d}r_2\,.
\end{aligned}
\end{equation}
By applying the same fitting procedure used above we obtain
\begin{equation}
\begin{aligned}
\mathcal{M}_{nl,|l-2|}^{\left( 2 \right)}
&=\frac{8\pi}{5}
\left[ \left( \frac{35l^5}{12}-\frac{1085l^4}{48}+\frac{415l^3}{8}-\frac{1495l^2}{48}-\frac{193l}{24}+\frac{7}{4} \right) n^6 \right.
\\&
+\left( \frac{5l^3}{4}+15l^2-\frac{745l}{12}+\frac{2135}{48} \right) n^8
\left.
+\left( \frac{541}{48}-\frac{14l}{3} \right) n^{10} \right] ,
\\
\mathcal{M}_{nl,l+2}^{\left( 2 \right)}&=\frac{8\pi}{5}
\left[ \left( -\frac{35l^5}{12}-\frac{595l^4}{16}-\frac{4115l^3}{24}-\frac{5625l^2}{16}-\frac{2519l}{8}-\frac{395}{4} \right) n^6 \right.
\\&
+\left( -\frac{5l^3}{4}+\frac{45l^2}{4}+\frac{265l}{3}+\frac{1925}{16} \right) n^8
\left.+\left( \frac{14l}{3}+\frac{255}{16} \right) n^{10} \right] ,
\\
\mathcal{M}_{nl,l}^{\left( 2 \right)}&=\frac{8\pi}{5}\left[ \left( -\frac{21l^4}{16}-\frac{21l^3}{8}-\frac{147l^2}{16}-\frac{63l}{8}+\frac{7}{4} \right) n^6+\left( -\frac{45l^2}{8}-\frac{45l}{8}+\frac{345}{16} \right) n^8+\frac{143n^{10}}{16} \right]\,.
\end{aligned}
\end{equation}
Inserting the angular parts in Eq.~(\ref{angular_part_scalar}) with $\mu=0$,
we have the scalar quadrupole polarizability denoted by $\alpha^{(2)}_s$
\begin{align}\label{quadrupole}
\alpha^{(2)}_s&=
\sum_{l'} \mathcal{M}^{(2)}_{nl,l'} \mathcal{A}^{(2)0}_{l,l'}
=
\mathcal{M}^{(2)}_{nl,l-2} \mathcal{A}^{(2)0}_{l,l-2}
+\mathcal{M}^{(2)}_{nl,l} \mathcal{A}^{(2)0}_{l,l}
+\mathcal{M}^{(2)}_{nl,l+2} \mathcal{A}^{(2)0}_{l,l+2}
\nonumber\\&
=
\left[ -\frac{399}{40}l^2\left( l+1 \right) ^2-\frac{1581}{20}l\left( l+1 \right) -\frac{79}{2} \right] n^6+\left[ 3l\left( l+1 \right) +\frac{385}{8} \right] n^8+\frac{51}{8}n^{10}.
\end{align}

In the octupole case, $l'$ takes $|l-3|$, $|l-1|$, $l+1$, $l+3$, and
${Q}^{(3)}_\mu=\sqrt{4\pi/7}~r^3 Y_{3\mu} (\theta, \phi)$. The radial integrals are given by
\begin{equation}
\begin{aligned}
\mathcal{M}_{nl,l-3}^{\left( 3 \right)}
&=\left[ \left(
-\frac{3605l^7}{256}
+\frac{44765l^6}{256}
-\frac{198485l^5}{256}
+\frac{381605l^4}{256}
\right.\right.
\\&
\left.~~~~~~
-\frac{153139l^3}{128}
+\frac{44233l^2}{128}
-\frac{4209l}{64}
+\frac{405}{32} \right) n^8
\\&
~+\left( \frac{2975l^5}{256}-\frac{68005l^4}{256}+\frac{504245l^3}{384}-\frac{273525l^2}{128}+\frac{406735l}{384}-\frac{27265}{128} \right) n^{10}
\\&
~+\left( \frac{19859l^3}{768}+\frac{3619l^2}{256}-\frac{433307l}{768}+\frac{185885}{256} \right) n^{12}
\\&
~\left. +\left( \frac{20805}{256}-\frac{6075l}{256} \right) n^{14} \right]
\times\frac{8\pi}{7}, \notag
\end{aligned}
\end{equation}

\begin{equation}
\begin{aligned}
\mathcal{M}_{nl,l-1}^{\left( 3 \right)}&=
\left[ \left( -\frac{135l^7}{256}+\frac{765l^6}{128}-\frac{1935l^5}{128}+\frac{45l^4}{32}+\frac{24161l^3}{256}-\frac{3909l^2}{128}-\frac{6699l}{64}+\frac{621}{32} \right) n^8 \right.
\\&
~+\left( -\frac{315l^5}{256}-\frac{165l^4}{128}+\frac{1015l^3}{8}-\frac{2915l^2}{16}-\frac{87465l}{256}+\frac{32395}{128} \right) n^{10}
\\&
~+\left( \frac{2219l^3}{256}-\frac{6461l^2}{128}-\frac{21777l}{128}+\frac{10829}{32} \right) n^{12}
\\&
~\left. +\left( \frac{7365}{128}-\frac{2025l}{256} \right) n^{14} \right]
\times \frac{8\pi}{7},\notag
\end{aligned}
\end{equation}

\begin{equation}
\begin{aligned}
\mathcal{M}_{nl,l+1}^{\left( 3 \right)}&=
\left[ \left( \frac{135l^7}{256}+\frac{2475l^6}{256}+\frac{15885l^5}{256}+\frac{47385l^4}{256}+\frac{6413l^3}{32}-\frac{1707l^2}{32}-\frac{3801l}{32}+\frac{355}{16} \right) n^8 \right.
\\&
~+\left( \frac{315l^5}{256}+\frac{1245l^4}{256}-\frac{15325l^3}{128}-\frac{71455l^2}{128}-\frac{12875l}{32}+\frac{2285}{8} \right) n^{10}
\\&
~+\left( -\frac{2219l^3}{256}-\frac{19579l^2}{256}+\frac{11053l}{256}+\frac{115045}{256} \right) n^{12}
\\&
~\left. +\left( \frac{2025l}{256}+\frac{16755}{256} \right) n^{14} \right]\times \frac{8\pi}{7},\notag
\end{aligned}
\end{equation}

\begin{align}
\mathcal{M}_{nl,l+3}^{\left( 3 \right)}&= \nonumber
\left[ \left(
\frac{3605l^7}{256}
+\frac{4375l^6}{16}
+\frac{135695l^5}{64}
+\frac{67865l^4}{8}
+\frac{4839023l^3}{256}
\right.\right.
\\&\left.~~~~~~\nonumber
+\frac{188405l^2}{8}
+\frac{490659l}{32}
+\frac{65205}{16} \right) n^8
\\&\nonumber
~+\left( -\frac{2975l^5}{256}-\frac{1295l^4}{4}-\frac{239225l^3}{96}-\frac{249165l^2}{32}-\frac{7981925l}{768}-\frac{159985}{32} \right) n^{10}
\\&\nonumber
~+\left( -\frac{19859l^3}{768}-\frac{1015l^2}{16}+\frac{98861l}{192}+\frac{40915}{32} \right) n^{12}
\\&
~\left. +\left( \frac{6075l}{256}+105 \right) n^{14} \right]\times\frac{8\pi}{7}.
\end{align}
The scalar octupole polarizability denoted by $\alpha^{(3)}_s$ is thus
\begin{align}\label{octupole}
\alpha^{(3)}_s
&=
\sum_{l'} \mathcal{M}^{(3)}_{nl,l'} \mathcal{A}^{(3)0}_{l,l'}
\nonumber \\&
=
\left[ \frac{10005}{224}l^3\left( l+1 \right) ^3+\frac{503775}{448}l^2\left( l+1 \right) ^2+\frac{94335}{32}l\left( l+1 \right) +\frac{9315}{8} \right] n^8
\nonumber \\&
~+\left[ -\frac{24855}{448}l^2\left( l+1 \right) ^2-\frac{18845}{14}l\left( l+1 \right) -\frac{22855}{16} \right] n^{10}
\nonumber \\&
~+\left[ \frac{5845}{16}-\frac{595}{32}l\left( l+1 \right) \right] n^{12}+30n^{14}
\end{align}
For higher-order polarizabilities, we simply list the results for the cases of $\lambda=4$ and $\lambda=5$
\begin{align}
\alpha _{s}^{\left( 4 \right)}&=
\left[ -\frac{2225}{12}l^4\left( l+1 \right) ^4-\frac{1152875}{96}l^3\left( l+1 \right) ^3
-\frac{10350025}{96}l^2\left( l+1 \right) ^2 \right.
\nonumber \\&\left.
~~~~~~-\frac{3003935}{16}l\left( l+1 \right) -\frac{124299}{2} \right] n^{10}
\nonumber \\&
~+\left[ \frac{9835}{24}l^3\left( l+1 \right) ^3+\frac{8539195}{384}l^2\left( l+1 \right) ^2+\frac{7460075}{64}l\left( l+1 \right) +\frac{2737085}{32} \right] n^{12}
\nonumber \\&
~+\left[ -\frac{18165}{128}l^2\left( L+1 \right) ^2-\frac{398055}{32}l\left( l+1 \right) -\frac{3012009}{128} \right] n^{14}
\nonumber \\&
~+\left[ \frac{137235}{64}-\frac{13545}{64}l\left( l+1 \right) \right] n^{16}
\nonumber \\&
~+\frac{16495}{128} n^{18},
\end{align}
\begin{equation}
\begin{aligned}
\alpha^{(5)}_s &=
\left[ \frac{525315}{704}l^5\left( l+1 \right) ^5+\frac{595596225}{5632}l^4\left( l+1 \right) ^4+\frac{3466632645}{1408}l^3\left( l+1 \right) ^3
\right.
\\&\left.
~~~~~~+\frac{18923161215}{1408}l^2\left( l+1 \right) ^2+\frac{6365941155}{352}l\left( l+1 \right) +\frac{42728175}{8} \right] n^{12}
\\&
~+\left[ -\frac{13401675}{5632}l^4\left( l+1 \right) ^4-\frac{192584175}{704}l^3\left( l+1 \right) ^3-\frac{6068288037}{1408}l^2\left( l+1 \right) ^2 \right.
\\&\left.
~~~~~~-\frac{868148127}{64}l\left( l+1 \right) -\frac{973745157}{128} \right] n^{14}
\\&
~+\left[ \frac{34965}{16}l^3\left( l+1 \right) ^3+\frac{170846235}{704}l^2\left( l+1 \right) ^2+\frac{6106752099}{2816}l\left( l+1 \right) +\frac{1316905205}{512} \right] n^{16}
\\&
~+\left[ \frac{224721}{704}l^2\left( l+1 \right) ^2-\frac{60183963}{704}l\left( l+1 \right) -\frac{134357685}{512} \right] n^{18}
\\&
~+\left[ \frac{5498955}{512}-\frac{359289}{256}l\left( l+1 \right) \right] n^{20}
\\&
~+\frac{272713}{512}n^{22}.
\end{aligned}
\end{equation}
As we assumed at the beginning of our fitting process, the scalar polarizabilities can be expressed as a polynomial of the principal quantum number $n$, written as $\sum_i C_i n^{x_i}$, where $C_i$ are polynomials of the orbital quantum number $l$, written as $C_i = \sum_j d_{ij} l^{y_{ij}}$. We can now see that the actual $l$ dependence of $C_i$ is a polynomial of $l(l+1)$.

\begin{table}[!h]
\caption{Quadrupole, octupole, and hexadecapole polarizabilities.}
\label{table.scalar.compared}
\begin{tabular}{ccc}
\hline\hline
  $Z$  & This work & Ref.~\cite{PhysRevA.86.012505} \\ \hline
  \multicolumn{3}{c}{$\alpha_s^{(2)}$}    \\
$1$ & $15$             & $14.998 829 822 856 441 699$              \\
$2$ & $0.234375$       & $0.234 301 867 935 791 210 0$             \\
$5$ & $0.00096‬$       & $9.581 285 372 324 045 392\times 10^{-4}$ \\ 
 \multicolumn{3}{c}{$\alpha_s^{(3)}$}    \\
$1$ & $131.25$         & $131.237 821 447 844 662$                 \\
$2$ & $0.5126953125$   & $0.512 505 037 523 770 47$                \\
$5$ & $0.000336$       & $3.352 210 608 787 016 2 \times 10^{-4}$  \\ 
 \multicolumn{3}{c}{$\alpha_s^{(4)}$}    \\
$1$ & $2126.25$        & $2126.028 674 499 128 83$                 \\
$2$ & $2.076416015625$ & $2.075 551 546 061 205 19$                \\
$5$ & $0.000217728$    & $2.171 618 426 945 541 1\times 10^{-4}$   \\ \hline\hline
\end{tabular}
\end{table}
We now compare our results with the calculations of Tang {\it et al.}~\cite{PhysRevA.86.012505}, as listed in Table~\ref{table.scalar.compared}, for
the cases of scalar quadrupole, octupole, and hexadecapole polarizabilities of the ground state hydrogen-like systems. It should be noted that
the work of Tang {\it et al.} is based on the Dirac-Coulomb Hamiltonian, which means that their results include relativistic corrections, whereas ours are fully nonrelativistic.

There is a scaling relation for multipole polarizabilities of different nuclear charge $Z$.
The multipole polarizability has a general expression $\left< \bm{r}^\lambda \frac{1}{E-H} \bm{r}^\lambda \right>$, where $\lambda$ is the multiplicity. By noting that $\bm{r}$ scales as $1/Z$ and $E-H$ scales as $Z^2$, the term $\left< \bm{r}^\lambda \frac{1}{E-H} \bm{r}^\lambda \right>$ thus scales as $1/Z^{2\lambda+2}$. We therefore have the following relation
\begin{equation}
\begin{aligned}
\alpha^{(\lambda)}_{Z}=\frac{\alpha^{(\lambda)}_{1}}{Z^{2\lambda+2}}\,,
\end{aligned}
\end{equation}
as reflected in Table~\ref{table.scalar.compared}.

As the multiplicity $\lambda$ increases, the range of the $n$-power index $x_i$ becomes larger,  $2\lambda+2i$, with $i=1,2,\cdots,\lambda+1$. Taking $\lambda=5$ as an example, the $n$-power index values are $12, 14, 16, 18, 20, 22$, which are exactly ranging from $2\lambda+2$ to $2\lambda+2(\lambda+1)$ with a step of 2. For a certain $C_i$, the $l(l+1)$-power index $y_{ij}$ ranges from 0 to $(x_{\mathrm{max}}-x_i)/2$. Taking $\lambda=5$ as an example again, $x_{\mathrm{max}}=22$, then the coefficient of $n^{12}$ has the terms from $[l(l+1)]^y$ in which $y$ ranges from 0 to $(22-12)/2=5$. The correspondent coefficient of $n^{14}$ has terms of $[l(l+1)]^0 \sim [l(l+1)]^4$ just because of $(22-14)/2=4$. The highest-order term $n^{22}$ has coefficient with only $[l(l+1)]^0$. This pattern can be tested in different $\lambda$ cases we listed. However, the pattern of $l(l+1)$ coefficients $d_{ij}$ has not been found, which may be an interesting question.

\section{application on blackbody radiation shift} \label{BBR-shift}

Blackbody radiation (BBR) shift in an atomic system is due to the AC Stark effect, which is related to the dynamical polarizability of the system.
The BBR is isotropic and thus it can be described by the scalar polarizability. Some studies on BBR shifts to atomic energy levels can be found in \cite{PhysRevA.78.042504, PhysRevA.23.2397, PhysRevD.28.340, PhysRevA.92.022508, zhou2017the}, including relativistic and QED corrections.
Here we only focus on the nonrelativistic energy shift due to the BBR, which is given by
\begin{equation}\label{BBR_energy_shift}
\begin{aligned}
\Delta E =
\frac{-2e^2}{\pi}
\int{\text{d}\omega \frac{\omega^3}{e^{\omega /(k_B T)}-1}}\left< \boldsymbol{r}\frac{\left( E_0-H_0 \right)}{\left( E_0-H_0 \right) ^2-\omega ^2}\boldsymbol{r} \right>,
\end{aligned}
\end{equation}
where $[(E_0-H_0)^2 - \omega^2]^{-1}$ is the Green function, $\omega$ is the photon energy from BBR, and $k_B$ is the Boltzmann constant. Usually the photon energy is much smaller than the energy intervals in atom so that we may expand $[(E_0-H_0)^2 - \omega^2]^{-1}$ according to
$$
\frac{1}{\left( E_0-H_0 \right) ^2-\omega ^2}\sim \frac{1}{\left( E_0-H_0 \right) ^2}+\frac{\omega ^2}{\left( E_0-H_0 \right) ^4}+\frac{\omega ^4}{\left( E_0-H_0 \right) ^6}+\cdots\,.
$$
Then we recast Eq.~\eqref{BBR_energy_shift} into the following form
\begin{equation}\label{energy_shift_approx}
\begin{aligned}
\Delta E&=
\frac{-2e^2}{\pi}
\int_0^{\infty}{\frac{\omega^3}{e^{\omega /k_BT}-1}}\left< \bm r\left[ \frac{1}{\left( E_0-H_0 \right)}+\frac{\omega ^2}{\left( E_0-H_0 \right) ^3}+\cdots \right] \bm r \right> \text{d}\omega
\\&
=
\Delta E^{(1)}
+\Delta E^{(2)}
+\cdots \,,
\end{aligned}
\end{equation}
where the $(E_0-H_0)^{-1}$ is the reduced Green function shown in Eqs.~\eqref{GF_general} and \eqref{Radial_GF}.
We can see that the BBR shift can be expressed in terms of static polarizabilities.
The leading-order BBR shift is
\begin{equation}\label{E1}
\begin{aligned}
\Delta E^{\left( 1 \right)}=
\frac{-2e^2}{\pi}
\int_0^{\infty}{\text{d}\omega \frac{\omega ^3}{e^{\omega /k_BT}-1}}\left< \bm r\frac{1}{E_0-H_0} \bm r \right>\,,
\end{aligned}
\end{equation}
in natural units, or
\begin{equation}\label{E1_au}
\begin{aligned}
\Delta E^{(1)}
=
\frac{-e^2}{2}\frac{\hbar}{\pi ^2c^3\epsilon _0}\int_0^{\infty}{}\text{d}\omega \frac{\omega ^3}{e^{\hbar\omega /k_BT}-1}\left< \boldsymbol{r}\frac{1}{\left( E_0-H_0 \right)}\boldsymbol{r} \right> \frac{a_{0}^{2}}{E_h},
\end{aligned}
\end{equation}
in SI unites,
where $c$ is the speed of light, $\epsilon_0$ is the permittivity, $a_0$ is Bohr radius, and $E_\mathrm{h}$ is the Hartree energy. In the above equation, $\left<\bm{r} (E_0-H_0)^{-1} \bm{r}\right>$ is nothing but the static dipole polarizability, as shown in Eq.~(\ref{eq_alpha1}).
The remaining integration over $\omega$ can be done and the result is
\begin{equation}
\begin{aligned}
\int_0^\infty \mathrm{d}\omega \frac{\omega^3}{e^{\hbar\omega/k_B T}-1}
=
\frac{\pi ^4 k_B^4 T^4}{15 \hbar^4}.
\end{aligned}
\end{equation}

In Table \ref{table1}, we list the BBR shifts to the $1S$ and $2S$ states of hydrogen. Our results for the $1S$ state are in good agreements with Ref.~\cite{PhysRevA.78.042504}. For the $2S$ state, however, there exist significant discrepancies between our results and Ref.~\cite{PhysRevA.78.042504}, especially at low temperatures. We will discuss this issue later.
\begin{table*}
\centering
\caption{
Blackbody radiation shifts to the $1S$ and $2S$ states of hydrogen, in Hz.
}
\label{table1}
\begin{tabular}{cccc}
\hline
\hline
Temperature~(K) & This work & Ref.~\cite{PhysRevA.78.042504} & Ref.~\cite{PhysRevA.23.2397}
\\
\hline
  \multicolumn{4}{c}{$1S$}    \\
300 & $-3.8786\times 10^{-2}$ & $-3.88\times 10^{-2}$ & $-0.04128$
\\
77 & $-1.6832\times 10^{-4}$  & $-1.68\times 10^{-4}$
\\
3 & $-1.2258\times 10^{-9}$ & $-1.22 \times 10^{-9}$
\\
  \multicolumn{4}{c}{$2S$}    \\
  300 & $-1.0343$ & $-9.89\times 10^{-1}$ & $-1.077$
\\
77 & $-4.4887 \times 10^{-3}$  & $-1.44\times 10^{-3}$
\\
3 & $-3.2689 \times 10^{-8}$ & $7.79\times 10^{-7}$
\\
\hline
\hline
\end{tabular}
\end{table*}

The second-order BBR shift is given by
\begin{equation}
\begin{aligned}
\Delta E^{(2)}
=
\frac{-e^2}{2}\frac{\hbar}{\pi ^2c^3\epsilon _0}\int_0^{\infty}{}\text{d}\omega \frac{\omega ^5}{e^{\hbar\omega /k_BT}-1}\left< \boldsymbol{r}\frac{1}{\left( E_0-H_0 \right)^3}\boldsymbol{r} \right> \frac{a_{0}^{2}}{E^3_h}\,.
\end{aligned}
\end{equation}
In the coordinate representation, the key quantity we need to calculate is
$$
\mathcal{I}^{(3)}_{nll'}\equiv
\int{}\psi _{nlm}^{\dag}\left( \boldsymbol{r}_1 \right) \boldsymbol{r}_1G_{1}^{'}\left( \boldsymbol{r}_1,\boldsymbol{r}_2;E_n \right) G_{2}^{'}\left( \boldsymbol{r}_2,\boldsymbol{r}_3;E_n \right) G_{3}^{'}\left( \boldsymbol{r}_3,\boldsymbol{r}_4;E_n \right) \boldsymbol{r}_4\psi _{nlm}\left( \boldsymbol{r}_4 \right) \text{d}\boldsymbol{r}_1\text{d}\boldsymbol{r}_2\text{d}\boldsymbol{r}_3\text{d}\boldsymbol{r}_4.
$$
The angular part of this integral can be easily evaluated due to the orthogonality relation of spherical harmonics. The radial part can be written
in the form
$$
\int{r_{1}^{2}\text{d}r_1r_{2}^{2}\text{d}r_2r_{3}^{2}\text{d}r_3r_{4}^{2}\text{d}r_4}R_{nl}\left( r_1 \right) g_{l_1}\left( r_1,r_2;E_n \right) g_{l_2}\left( r_2,r_3;E_n \right) g_{l_3}\left( r_3,r_4;E_n \right) R_{nl}\left( r_4 \right).
$$
Since the angular part contains $\delta_{l_1 l_2}$ and $\delta_{l_2,l_3}$, the three radial functions $g_{l_i}$ will have the same intermediate orbital angular quantum numbers, $i.e.$, $|l-1|$ and $l+1$.
By applying the same method introduced in Sec.~\ref{Sec.Method}, we can obtain the following expressions
\begin{equation}
\begin{aligned}
\mathcal{M}_{nl,|l-1|}^{(1,3)}=&
\left(-\frac{5 l^5}{96}+\frac{45 l^4}{64}-\frac{163 l^3}{48}+\frac{495 l^2}{64}-\frac{647 l}{96}
+\frac{21}{16}\right) n^8
\\&
+\left(-\frac{35 l^3}{96}+\frac{55 l^2}{16}-\frac{845 l}{96}+\frac{341}{64}\right) n^{10}
+\left(\frac{91}{64}-\frac{7 l}{12}\right) n^{12},
\notag
\end{aligned}
\end{equation}

\begin{align}
\mathcal{M}_{nl,l+1}^{(1,3)}
&=
\left(
\frac{5 l^5}{96}
+\frac{185 l^4}{192}
+\frac{323 l^3}{48}
+\frac{4351 l^2}{192}
+\frac{1135 L}{32}
+\frac{319}{16}
\right) n^8
\nonumber\\&
+\left(
\frac{35 l^3}{96}
+\frac{145 l^2}{32}
+\frac{805 l}{48}
+\frac{3443}{192}
\right) n^{10}
+\left(\frac{7 l}{12}+\frac{385}{192}\right) n^{12}.
\end{align}
The corresponding angular parts are
\begin{equation}
\begin{aligned}
\mathcal{A}^{(1,3)}_{l,|l-1|}=\frac{l}{6 l+3},~
\mathcal{A}^{(1,3)}_{l,l+1}=\frac{l+1}{6 l+3}.
\end{aligned}
\end{equation}
We thus have
\begin{equation}
\begin{aligned}
\mathcal{I}^{(3)}_{nl}&= \sum_{l'} \mathcal{I}^{(3)}_{nll'}=
\mathcal{M}_{nl,l+1}^{(1,3)} \mathcal{A}^{(1,3)}_{l,l+1}
+
\mathcal{M}_{nl,|l-1|}^{(1,3)} \mathcal{A}^{(1,3)}_{l,|l-1|}
\\&
=
\left[\frac{55}{96} l^2 (l+1)^2
+\frac{539}{48} l (l+1)
+\frac{319}{24}\right] n^8
+\left[\frac{25}{9} l (l+1)
+\frac{3443}{288}\right] n^{10}
+\frac{385 n^{12}}{288}\,.
\end{aligned}
\end{equation}
The corresponding BBR shift to the $1S$ state of hydrogen at $300~\text K$ is $-3.8878\times 10^{-6}~\mathrm{Hz}$, which is much smaller than the leading-order one and can be omitted in most cases.

In Appendix~\ref{appx_Integral}, we give the analytical expressions for $\langle \bm{r} (E_0-H_0)^{-2} \bm{r}\rangle$ and $\left< \bm{r} (E_0-H_0)^{-4} \bm{r}\right>$, which will be useful in higher-order perturbation calculations.
For instance, one should calculate $\left< \bm{r} (E_0-H_0)^{-2} \bm{r}\right>$ when considering the second-order perturbation of Eq.~\eqref{E1_au}.

Our approach here is nonrelativistic in nature, which works well for the ground states, as shown in Table~\ref{table1}.
However, for exited $nS$ states, even for $n=2$, the discrepancy appears, especially at lower temperatures. In Ref.~\cite{PhysRevA.78.042504}, the radial wave function is treated under the Schr\"odinger-Pauli approximation, whereas the angular part is treated using a Dirac spinor and the Dirac angular quantum number. Further, they also consider the Lamb shift and fine-structure contributions. Table~\ref{table1} also shows that, for the
 $2S$ state, the discrepancy between our results and Ref.~\citep{PhysRevA.78.042504} becomes more significant as the temperature decreases, due to the relativistic effect. In the nonrelativistic limit, the BBR shift acts as $T^4/Z^4$; whereas the relativistic correction acts as $(Z\alpha T)^2$. Therefore, the relativistic correction becomes more important for high $Z$ and/or low $T$~\cite{PhysRevA.78.042504, zhou2017the}.


\section{Summary}
In this work, based on the analytical wave function and the reduced Green function, we have obtained the analytical expressions for the scalar multipole polarizabilities of hydrogen-like ions through a numerical fitting procedure. These analytical expressions can be expressed as a polynomial of $n$ and $l(l+1)$. We have applied our results to the BBR calculations and found that the relativistic effects are particularly important at low temperatures. Our analytical formulas can be served as a benchmark for other computational methods.

\section{Acknowledgement}
The authors wish to thank Dr. Xiaofeng Wang and Dr. Liyan Tang for helpful discussions, and Prof. Zong-Chao Yan for valuable suggestions. This work was supported by the National Natural Science Foundation of China (No. 11674253). X.-S. M. was also supported by the National Natural Science Foundation of China (Nos. 11474316 and 91636216) and the Strategic Priority Research Program of the Chinese Academy of Sciences (No. XDB21020200).

\newpage

\begin{appendix}
\section{Supplementary materials}

\subsection{A table on fitting process}
\label{Ftable}

\begin{table}[!htb]
\tiny
\caption{An example of the $c_i$ fitting. The values of radial integrals in different $n$ are also listed. The coefficients $c_i$ have only non-zero $c_1$ and $c_2$. For simplicity, we only list the results of $l=0,1,9$ for $l'=l+1$, and $l=1,2,9$ for $l'=|l-1|$.}
\label{fitting_table}
\begin{tabular}{cccc}
\hline \hline
\multicolumn{4}{c}{$l'=l+1$}                                                                                                                                                                                                                                                                                                                                                                                                                                                                                                                                                                                                                                                                                                                                                                                                                                                              \\ \hline
$$                                                             & \multicolumn{1}{c}{$l=0$}                                                                                                                                                                                                                       & \multicolumn{1}{c}{$l=1$}                                                                                                                                                                                                                   & \multicolumn{1}{c}{$l=9$}                                                                                                                                                                                                                                                                        \\
\begin{tabular}[c]{@{}c@{}}
$\mathcal M^{(1)}(n,l)$\\ $n=1,2,\cdots,20$\end{tabular} & \begin{tabular}[c]{@{}l@{}}
$$\{27/4,180,6075/4,7488,\\ 106875/4,76788,756315/4,\\ 414720,3326427/4,\\ 1552500,10936827/4,\\ 4587840,29560635/4,\\ 11495988,69406875/4,\\ 25509888,146579355/4,\\ 51569460,285012027/4,\\ 96840000\}
$$
\end{tabular} &
\begin{tabular}[c]{@{}l@{}}
$$
\{390,2835,12768,\\ 43125,119718,288120,\\ 622080,1233468,2283750,\\ 3996993,6674400,\\ 10710375,16610118,\\ 25008750,36691968,\\ 52618230,73942470,\\ 102041343,138540000,\\ 185340393\}
$$
\end{tabular}          & \begin{tabular}[c]{@{}l@{}}
$$
\{12993750,21214809,\\ 33444576,51152751,\\ 76193334,110868750,\\ 157999104,220996566,\\ 303944886,411684039,\\ 549900000,725219649,\\ 945310806,1218987396,\\ 1556319744,1968750000,\\ 2469212694,3072260421,\\ 3794194656,4653201699\}
$$
\end{tabular}                \\
\begin{tabular}[c]{@{}c@{}}$c_i^{(1)}$ \\ $(i=1,\cdots,20)$\end{tabular}
& \multicolumn{1}{c}
{$\{3/2,21/4,0,0,\cdots,0\}$}                                                                                                                                                                                              &
$\{17/8,127/8,0,0,\cdots,0\}                                                                                                                                               $
& $\{57/8,4695/8,0,0,\cdots,0\}$                                                                                                                                                                                                                                                                     \\ \hline
\multicolumn{4}{c}{$l'=|l-1|$}                                                                                                                                                                                                                                                                                                                                                                                                                                                                                                                                                                                                                                                                                                                                                                                                                                                              \\ \hline
                                                             & \multicolumn{1}{c}{$l=1$}                                                                                                                                                                                                                       & \multicolumn{1}{c}{$l=2$}                                                                                                                                                                                                                                      & \multicolumn{1}{c}{$l=9$}                                                                                                                                                                                                                                                                       \\
\begin{tabular}[c]{@{}c@{}}
$\mathcal M^{(1)}(n,l)$ \\ $n=1,2,\cdots,20$ \end{tabular} & \begin{tabular}[c]{@{}l@{}}
$$
\{12,162,960,3750,\\ 11340,28812,64512,\\ 131220,247500,439230,\\ 741312,1199562,\\ 1872780,2835000,\\ 4177920,6013512,\\ 8476812,11728890,\\ 15960000,21392910\}
$$
\end{tabular}                                      & \begin{tabular}[c]{@{}l@{}}
$$
\{-243,-1440,-5625,\\ -17010,-43218,-96768,\\ -196830,-371250,\\ -658845,-1111968,\\ -1799343,-2809170,\\ -4252500,-6266880,\\ -9020268,-12715218,\\ -17593335,-23940000,\\ -32089365,-42429618\}
$$
\end{tabular} &
\begin{tabular}[c]{@{}l@{}}
$$
\{-6172500,-10497597,\\ -17133120,-26990145,\\ -41229972,-61306875,\\ -89014272,-126534315,\\ -176490900,-242006097,\\ -326760000,-435053997,\\ -571877460,-742977855,\\ -954934272,-1215234375,\\ -1532354772,-1915844805,\\ -2376413760,-2926021497\}
$$
\end{tabular} \\
\begin{tabular}[c]{@{}c@{}} $c_i^{(1)}$ \\ $(i=1,\cdots,20)$\end{tabular}  & $\{1/4,-1/4,0,0,\cdots,0\}$                                                                                                                                                                                                                     & $\{-3/8,3/8,0,0,\cdots,0\}                                  $                                        &
$\{-19/4,-569/4,0,\cdots,0\}                                                                                                                                                                                                                                                               $ \\ \hline\hline
\end{tabular}
\end{table}

\subsection{Integrals $\left<\bm{r}G^{\prime 2}\bm{r}\right>$ and $\left<\bm{r}G^{\prime 4}\bm{r}\right>$} \label{appx_Integral}

Here we present the analytical results for the integrals $\left<\bm{r}G^{\prime 2}\bm{r}\right>$ and $\left<\bm{r}G^{\prime 4}\bm{r}\right>$. In the coordinate representation, they can be written as
\begin{equation}
\begin{aligned}
\mathcal{I}_{n,ll'}^{\left( 2 \right)}\equiv \int{}\psi _{nlm}^{\dag}\left( \boldsymbol{r}_1 \right) \boldsymbol{r}_1G_{1}^{'}\left( \boldsymbol{r}_1,\boldsymbol{r}_2;E_0 \right) G_{2}^{'}\left( \boldsymbol{r}_2,\boldsymbol{r}_3;E_0 \right) \boldsymbol{r}_3\psi _{nlm}\left( \boldsymbol{r}_3 \right) \text{d}\boldsymbol{r}_1\text{d}\boldsymbol{r}_2\text{d}\boldsymbol{r}_3,
\end{aligned}
\end{equation}
\begin{equation}
\begin{aligned}
\mathcal{I}_{n,ll'}^{\left( 4 \right)}\equiv &
\int{}\psi _{nlm}^{\dag}\left( \boldsymbol{r}_1 \right) \boldsymbol{r}_1
G_{1}^{'}\left( \boldsymbol{r}_1,\boldsymbol{r}_2;E_0 \right)
G_{2}^{'}\left( \boldsymbol{r}_2,\boldsymbol{r}_3;E_0 \right)
\\&
\times
G_{2}^{'}\left( \boldsymbol{r}_3,\boldsymbol{r}_4;E_0 \right)
G_{2}^{'}\left( \boldsymbol{r}_4,\boldsymbol{r}_5;E_0 \right) \boldsymbol{r}_5
\psi _{nlm}\left( \boldsymbol{r}_5 \right) \text{d}\boldsymbol{r}_1\text{d}\boldsymbol{r}_2\text{d}\boldsymbol{r}_3\text{d}\boldsymbol{r}_4\text{d}\boldsymbol{r}_5.
\end{aligned}
\end{equation}
Using Eqs.~\eqref{Radial_WF}, \eqref{Radial_GF}, and ~\eqref{radial_general_integral} and applying the fitting procedure, we can obtain the following analytical expressions, after summing over $l'$
\begin{equation}
\begin{aligned}
\mathcal{I}_{nl}^{\left( 2 \right)}=
\left[\frac{5}{48} l^2(l+1)^2 +\frac{14}{3} l(l+1)+\frac{163}{24}\right] n^6
+\left[\frac{5}{12} (l+1) l+\frac{61}{16}\right] n^8
+\frac{7 n^{10}}{48},
\end{aligned}
\end{equation}

\begin{equation}
\begin{aligned}
\mathcal{I}_{nl}^{(4)}=&
\left[ \frac{35}{3456}l^3\left( l+1 \right) ^3+\frac{3595}{1728}l^2\left( l+1 \right) ^2+\frac{7385}{288}l\left( l+1 \right) +\frac{1255}{48} \right] n^{10}
\\&
+\left[ \frac{385}{3456}l^2\left( l+1 \right) ^2+\frac{40765}{3456}l\left( l+1 \right) +\frac{58145}{1728} \right] n^{12}
\\&
+\left[ \frac{1393}{3456}l\left( l+1 \right) +\frac{2779}{384} \right] n^{14}+\frac{491n^{16}}{3456}.
\end{aligned}
\end{equation}
\end{appendix}

\bibliography{SPHydrogen}

\begin{thebibliography}{17}%
\makeatletter
\providecommand \@ifxundefined [1]{%
 \@ifx{#1\undefined}
}%
\providecommand \@ifnum [1]{%
 \ifnum #1\expandafter \@firstoftwo
 \else \expandafter \@secondoftwo
 \fi
}%
\providecommand \@ifx [1]{%
 \ifx #1\expandafter \@firstoftwo
 \else \expandafter \@secondoftwo
 \fi
}%
\providecommand \natexlab [1]{#1}%
\providecommand \enquote  [1]{``#1''}%
\providecommand \bibnamefont  [1]{#1}%
\providecommand \bibfnamefont [1]{#1}%
\providecommand \citenamefont [1]{#1}%
\providecommand \href@noop [0]{\@secondoftwo}%
\providecommand \href [0]{\begingroup \@sanitize@url \@href}%
\providecommand \@href[1]{\@@startlink{#1}\@@href}%
\providecommand \@@href[1]{\endgroup#1\@@endlink}%
\providecommand \@sanitize@url [0]{\catcode `\\12\catcode `\$12\catcode
  `\&12\catcode `\#12\catcode `\^12\catcode `\_12\catcode `\%12\relax}%
\providecommand \@@startlink[1]{}%
\providecommand \@@endlink[0]{}%
\providecommand \url  [0]{\begingroup\@sanitize@url \@url }%
\providecommand \@url [1]{\endgroup\@href {#1}{\urlprefix }}%
\providecommand \urlprefix  [0]{URL }%
\providecommand \Eprint [0]{\href }%
\providecommand \doibase [0]{http://dx.doi.org/}%
\providecommand \selectlanguage [0]{\@gobble}%
\providecommand \bibinfo  [0]{\@secondoftwo}%
\providecommand \bibfield  [0]{\@secondoftwo}%
\providecommand \translation [1]{[#1]}%
\providecommand \BibitemOpen [0]{}%
\providecommand \bibitemStop [0]{}%
\providecommand \bibitemNoStop [0]{.\EOS\space}%
\providecommand \EOS [0]{\spacefactor3000\relax}%
\providecommand \BibitemShut  [1]{\csname bibitem#1\endcsname}%
\let\auto@bib@innerbib\@empty
\bibitem [{\citenamefont {Yan}(2000)}]{PhysRevA.62.052502}%
  \BibitemOpen
  \bibfield  {author} {\bibinfo {author} {\bibfnamefont {Z.-C.}\ \bibnamefont
  {Yan}},\ }\href {\doibase 10.1103/PhysRevA.62.052502} {\bibfield  {journal}
  {\bibinfo  {journal} {Phys. Rev. A}\ }\textbf {\bibinfo {volume} {62}},\
  \bibinfo {pages} {052502} (\bibinfo {year} {2000})}\BibitemShut {NoStop}%
\bibitem [{\citenamefont {Bhatti}\ \emph {et~al.}(2003)\citenamefont {Bhatti},
  \citenamefont {Coleman},\ and\ \citenamefont {Perger}}]{PhysRevA.68.044503}%
  \BibitemOpen
  \bibfield  {author} {\bibinfo {author} {\bibfnamefont {M.~I.}\ \bibnamefont
  {Bhatti}}, \bibinfo {author} {\bibfnamefont {K.~D.}\ \bibnamefont {Coleman}},
  \ and\ \bibinfo {author} {\bibfnamefont {W.~F.}\ \bibnamefont {Perger}},\
  }\href {\doibase 10.1103/PhysRevA.68.044503} {\bibfield  {journal} {\bibinfo
  {journal} {Phys. Rev. A}\ }\textbf {\bibinfo {volume} {68}},\ \bibinfo
  {pages} {044503} (\bibinfo {year} {2003})}\BibitemShut {NoStop}%
\bibitem [{\citenamefont {Cohen}\ and\ \citenamefont
  {Themelis}(2006)}]{cohen2006numerical}%
  \BibitemOpen
  \bibfield  {author} {\bibinfo {author} {\bibfnamefont {S.}~\bibnamefont
  {Cohen}}\ and\ \bibinfo {author} {\bibfnamefont {S.~I.}\ \bibnamefont
  {Themelis}},\ }\href@noop {} {\bibfield  {journal} {\bibinfo  {journal} {J.
  Chem. Phys.}\ }\textbf {\bibinfo {volume} {124}},\ \bibinfo {pages} {134106}
  (\bibinfo {year} {2006})}\BibitemShut {NoStop}%
\bibitem [{\citenamefont {Tang}\ \emph {et~al.}(2012)\citenamefont {Tang},
  \citenamefont {Zhang}, \citenamefont {Zhang}, \citenamefont {Jiang},\ and\
  \citenamefont {Mitroy}}]{PhysRevA.86.012505}%
  \BibitemOpen
  \bibfield  {author} {\bibinfo {author} {\bibfnamefont {L.-Y.}\ \bibnamefont
  {Tang}}, \bibinfo {author} {\bibfnamefont {Y.-H.}\ \bibnamefont {Zhang}},
  \bibinfo {author} {\bibfnamefont {X.-Z.}\ \bibnamefont {Zhang}}, \bibinfo
  {author} {\bibfnamefont {J.}~\bibnamefont {Jiang}}, \ and\ \bibinfo {author}
  {\bibfnamefont {J.}~\bibnamefont {Mitroy}},\ }\href {\doibase
  10.1103/PhysRevA.86.012505} {\bibfield  {journal} {\bibinfo  {journal} {Phys.
  Rev. A}\ }\textbf {\bibinfo {volume} {86}},\ \bibinfo {pages} {012505}
  (\bibinfo {year} {2012})}\BibitemShut {NoStop}%
\bibitem [{\citenamefont {Swainson}\ and\ \citenamefont
  {Drake}(1991{\natexlab{a}})}]{Swainson_1991a}%
  \BibitemOpen
  \bibfield  {author} {\bibinfo {author} {\bibfnamefont {R.~A.}\ \bibnamefont
  {Swainson}}\ and\ \bibinfo {author} {\bibfnamefont {G.~W.~F.}\ \bibnamefont
  {Drake}},\ }\href {\doibase 10.1088/0305-4470/24/1/019} {\bibfield  {journal}
  {\bibinfo  {journal} {J. Phys. A}\ }\textbf {\bibinfo {volume} {24}},\
  \bibinfo {pages} {79} (\bibinfo {year} {1991}{\natexlab{a}})}\BibitemShut
  {NoStop}%
\bibitem [{\citenamefont {Swainson}\ and\ \citenamefont
  {Drake}(1991{\natexlab{b}})}]{Swainson_1991b}%
  \BibitemOpen
  \bibfield  {author} {\bibinfo {author} {\bibfnamefont {R.~A.}\ \bibnamefont
  {Swainson}}\ and\ \bibinfo {author} {\bibfnamefont {G.~W.~F.}\ \bibnamefont
  {Drake}},\ }\href {\doibase 10.1088/0305-4470/24/1/020} {\bibfield  {journal}
  {\bibinfo  {journal} {J. Phys. A}\ }\textbf {\bibinfo {volume} {24}},\
  \bibinfo {pages} {95} (\bibinfo {year} {1991}{\natexlab{b}})}\BibitemShut
  {NoStop}%
\bibitem [{\citenamefont {Swainson}\ and\ \citenamefont
  {Drake}(1991{\natexlab{c}})}]{Swainson_1991c}%
  \BibitemOpen
  \bibfield  {author} {\bibinfo {author} {\bibfnamefont {R.~A.}\ \bibnamefont
  {Swainson}}\ and\ \bibinfo {author} {\bibfnamefont {G.~W.~F.}\ \bibnamefont
  {Drake}},\ }\href {\doibase 10.1088/0305-4470/24/8/022} {\bibfield  {journal}
  {\bibinfo  {journal} {J. Phys. A}\ }\textbf {\bibinfo {volume} {24}},\
  \bibinfo {pages} {1801} (\bibinfo {year} {1991}{\natexlab{c}})}\BibitemShut
  {NoStop}%
\bibitem [{\citenamefont {Krylovetsky}\ \emph {et~al.}(2001)\citenamefont
  {Krylovetsky}, \citenamefont {Manakov},\ and\ \citenamefont
  {Marmo}}]{krylovetsky2001generalized}%
  \BibitemOpen
  \bibfield  {author} {\bibinfo {author} {\bibfnamefont {A.~A.}\ \bibnamefont
  {Krylovetsky}}, \bibinfo {author} {\bibfnamefont {N.~L.}\ \bibnamefont
  {Manakov}}, \ and\ \bibinfo {author} {\bibfnamefont {S.~I.}\ \bibnamefont
  {Marmo}},\ }\href@noop {} {\bibfield  {journal} {\bibinfo  {journal} {Journal
  of Experimental and Theoretical Physics}\ }\textbf {\bibinfo {volume} {92}},\
  \bibinfo {pages} {37} (\bibinfo {year} {2001})}\BibitemShut {NoStop}%
\bibitem [{\citenamefont {Baye}(2012)}]{PhysRevA.86.062514}%
  \BibitemOpen
  \bibfield  {author} {\bibinfo {author} {\bibfnamefont {D.}~\bibnamefont
  {Baye}},\ }\href {\doibase 10.1103/PhysRevA.86.062514} {\bibfield  {journal}
  {\bibinfo  {journal} {Phys. Rev. A}\ }\textbf {\bibinfo {volume} {86}},\
  \bibinfo {pages} {062514} (\bibinfo {year} {2012})}\BibitemShut {NoStop}%
\bibitem [{\citenamefont {Porsev}\ and\ \citenamefont
  {Derevianko}(2006)}]{PhysRevA.74.020502}%
  \BibitemOpen
  \bibfield  {author} {\bibinfo {author} {\bibfnamefont {S.~G.}\ \bibnamefont
  {Porsev}}\ and\ \bibinfo {author} {\bibfnamefont {A.}~\bibnamefont
  {Derevianko}},\ }\href {\doibase 10.1103/PhysRevA.74.020502} {\bibfield
  {journal} {\bibinfo  {journal} {Phys. Rev. A}\ }\textbf {\bibinfo {volume}
  {74}},\ \bibinfo {pages} {020502} (\bibinfo {year} {2006})}\BibitemShut
  {NoStop}%
\bibitem [{\citenamefont {Jentschura}\ and\ \citenamefont
  {Haas}(2008)}]{PhysRevA.78.042504}%
  \BibitemOpen
  \bibfield  {author} {\bibinfo {author} {\bibfnamefont {U.~D.}\ \bibnamefont
  {Jentschura}}\ and\ \bibinfo {author} {\bibfnamefont {M.}~\bibnamefont
  {Haas}},\ }\href {\doibase 10.1103/PhysRevA.78.042504} {\bibfield  {journal}
  {\bibinfo  {journal} {Phys. Rev. A}\ }\textbf {\bibinfo {volume} {78}},\
  \bibinfo {pages} {042504} (\bibinfo {year} {2008})}\BibitemShut {NoStop}%
\bibitem [{\citenamefont {Buckingham}(1967)}]{buckingham1967permanent}%
  \BibitemOpen
  \bibfield  {author} {\bibinfo {author} {\bibfnamefont {A.~D.}\ \bibnamefont
  {Buckingham}},\ }\href@noop {} {\bibfield  {journal} {\bibinfo  {journal}
  {Advances in Chemical Physics: Intermolecular Forces}\ }\textbf {\bibinfo
  {volume} {12}},\ \bibinfo {pages} {107} (\bibinfo {year} {1967})}\BibitemShut
  {NoStop}%
\bibitem [{\citenamefont {Wang}\ and\ \citenamefont
  {Guo}(2000)}]{Introduction.to.Special.Functions}%
  \BibitemOpen
  \bibfield  {author} {\bibinfo {author} {\bibfnamefont {Z.-X.}\ \bibnamefont
  {Wang}}\ and\ \bibinfo {author} {\bibfnamefont {D.-R.}\ \bibnamefont {Guo}},\
  }\href {https://books.google.co.jp/books?id=3acajwEACAAJ} {\emph {\bibinfo
  {title} {Introduction to Special Functions (Chinese edition)}}}\ (\bibinfo
  {publisher} {Peking University Press},\ \bibinfo {year} {2000})\BibitemShut
  {NoStop}%
\bibitem [{\citenamefont {Farley}\ and\ \citenamefont
  {Wing}(1981)}]{PhysRevA.23.2397}%
  \BibitemOpen
  \bibfield  {author} {\bibinfo {author} {\bibfnamefont {J.~W.}\ \bibnamefont
  {Farley}}\ and\ \bibinfo {author} {\bibfnamefont {W.~H.}\ \bibnamefont
  {Wing}},\ }\href {\doibase 10.1103/PhysRevA.23.2397} {\bibfield  {journal}
  {\bibinfo  {journal} {Phys. Rev. A}\ }\textbf {\bibinfo {volume} {23}},\
  \bibinfo {pages} {2397} (\bibinfo {year} {1981})}\BibitemShut {NoStop}%
\bibitem [{\citenamefont {Donoghue}\ and\ \citenamefont
  {Holstein}(1983)}]{PhysRevD.28.340}%
  \BibitemOpen
  \bibfield  {author} {\bibinfo {author} {\bibfnamefont {J.~F.}\ \bibnamefont
  {Donoghue}}\ and\ \bibinfo {author} {\bibfnamefont {B.~R.}\ \bibnamefont
  {Holstein}},\ }\href {\doibase 10.1103/PhysRevD.28.340} {\bibfield  {journal}
  {\bibinfo  {journal} {Phys. Rev. D}\ }\textbf {\bibinfo {volume} {28}},\
  \bibinfo {pages} {340} (\bibinfo {year} {1983})}\BibitemShut {NoStop}%
\bibitem [{\citenamefont {Solovyev}\ \emph {et~al.}(2015)\citenamefont
  {Solovyev}, \citenamefont {Labzowsky},\ and\ \citenamefont
  {Plunien}}]{PhysRevA.92.022508}%
  \BibitemOpen
  \bibfield  {author} {\bibinfo {author} {\bibfnamefont {D.}~\bibnamefont
  {Solovyev}}, \bibinfo {author} {\bibfnamefont {L.}~\bibnamefont {Labzowsky}},
  \ and\ \bibinfo {author} {\bibfnamefont {G.}~\bibnamefont {Plunien}},\ }\href
  {\doibase 10.1103/PhysRevA.92.022508} {\bibfield  {journal} {\bibinfo
  {journal} {Phys. Rev. A}\ }\textbf {\bibinfo {volume} {92}},\ \bibinfo
  {pages} {022508} (\bibinfo {year} {2015})}\BibitemShut {NoStop}%
\bibitem [{\citenamefont {Zhou}\ \emph {et~al.}(2017)\citenamefont {Zhou},
  \citenamefont {Mei},\ and\ \citenamefont {Qiao}}]{zhou2017the}%
  \BibitemOpen
  \bibfield  {author} {\bibinfo {author} {\bibfnamefont {W.-P.}\ \bibnamefont
  {Zhou}}, \bibinfo {author} {\bibfnamefont {X.-S.}\ \bibnamefont {Mei}}, \
  and\ \bibinfo {author} {\bibfnamefont {H.-X.}\ \bibnamefont {Qiao}},\
  }\href@noop {} {\bibfield  {journal} {\bibinfo  {journal} {J. Phys. B}\
  }\textbf {\bibinfo {volume} {50}},\ \bibinfo {pages} {105001} (\bibinfo
  {year} {2017})}\BibitemShut {NoStop}%
\end{thebibliography}%
\end{document}